**9/11/12**

# Effect of resonant magnetic perturbations with toroidal mode numbers of 4 and 6 on ELMs in single null H-mode plasmas in MAST

A. Kirk, I.T. Chapman, J. Harrison, Yueqiang Liu, E. Nardon[1], S. Saarelma,
R. Scannell, A.J. Thornton and the MAST team

*EURATOM/CCFE Fusion Association, Culham Science Centre, Abingdon, Oxon OX14 3DB, UK.*
*[1]Association Euratom/CEA, CEA Cadarache, F-13108, St. Paul-lez-Durance, France*

## Abstract

The application of resonant magnetic perturbations (RMPs) with a toroidal mode number of n=4 or n=6 to lower single null plasmas in the MAST tokamak produces up to a factor of 5 increase in Edge Localized Mode (ELM) frequency and reduction in plasma energy loss associated with type-I ELMs.  A threshold current for ELM mitigation is observed above which the ELM frequency increases approximately linearly with current in the coils. Despite a large scan of parameters, complete ELM suppression has not been achieved. The results have been compared to modelling performed using either the vacuum approximation or including the plasma response.  During the ELM mitigated stage clear lobe structures are observed in visible-light imaging of the X-point region.   The size of these lobes is correlated with the increase in ELM frequency observed.   The characteristics of the mitigated ELMs are similar to those of the natural ELMs suggesting that they are type I ELMs which are triggered at a lower pressure gradient.  The application of the RMPs in the n=4 and n=6 configurations before the L-H transition has little effect on the power required to achieve H-mode while still allowing the first ELM to be mitigated.

## 1. Introduction

In order to avoid damage to in-vessel components in future devices, such as ITER, a mechanism to ameliorate the size of Edge Localized Modes (ELMs) is required [1]. One such amelioration mechanism relies on perturbing the magnetic field in the edge plasma region, either leading to more frequent smaller ELMs (ELM mitigation) or ELM suppression. This technique of Resonant Magnetic Perturbations (RMPs) has been employed to suppress type I ELMs at high collisionality on DIII-D [2] and ASDEX Upgrade [3] and at low collisionality on DIII-D [4]. The original interpretation of the low collisionality discharges on DIII-D was that the RMPs enhance the transport of particles or energy and keep the edge pressure gradient below the critical value that would trigger an ELM [4]. However, more recent finding suggest that the RMPs induce an island at the top of the pedestal and the transport due this island impedes the widening of the pedestal, which stops the peeling ballooning limit being reached [5]. However, at high collisionality on both DIII-D [2] and ASDEX Upgrade [3] the pedestal characteristics remain largely unchanged and the reason for the suppression of type I ELMs is unclear. In addition to complete suppression of the type I ELMs, both devices also can achieve periods of ELM mitigation. Similar periods of ELM mitigation have also been obtained on JET [6][7] and MAST [8].

The MAST ELM control system has been upgraded from two rows of 6 coils each [9] to a system of 18 coils (6 in the upper row and 12 in the lower row). These coils give considerably enhanced flexibility since they not only allow higher toroidal mode numbers (n=4 and n=6) but also allow an n=3 configuration with improved alignment of the magnetic perturbations with the plasma equilibrium, by allowing the pitch of the applied field to be varied during the shot. In addition, mixed spectra (e.g. n=3 and n=4) can be applied.

This paper presents results from the application of RMPs to Single Null Divertor (SND) discharges in MAST. Due to the up-down symmetry in the divertor coils on MAST, SND discharges are usually produced by shifting the plasma downwards. In this lower SND (LSND) magnetic configuration the plasma is far from the upper row of RMP coils and hence the perturbation is predominantly from the lower row of 12 coils. In this configuration resonant magnetic perturbations with toroidal mode numbers of n=3, 4 and 6 can be applied with similar strengths. For the n=3 configuration used in this paper, only 6 of the lower coils are used (i.e. the sign of the current in the coils is +0-0+0-0+0-0). The n=4 and n=6 configurations use all 12 lower coils with the sign of the currents in the coils being ++-++-++-++- and +-+-+-+-+-+- respectively. For the n=6 configuration the toroidal mode spectrum of the applied perturbation is effectively a pure n=6, the n=4 configuration has a sizeable n=8 side band. In all the cases considered in this paper no current is applied to the upper row of ELM coils. The poloidal cross section of the baseline scenario is shown in Figure 1.

The SND configuration has been chosen for this present study for three reasons. Firstly, the majority of ELM control work on other devices has been performed in a SND configuration and it is most relevant for future devices such as ITER. Secondly, the fact that the effective field from the ELM control coils falls off more rapidly in the higher toroidal mode number configurations favours minimising the distance between the plasma and the lower row of coils. This can most easily be achieved on MAST in a LSND plasma.

Finally the downward shift of the plasma required for the LSND configuration optimises the view of the X-point region.

The layout of this paper is as follows: Section 2 presents the general characteristics of the Lower SND plasma and the RMP configuration used together with the effects that the RMPs with different toroidal numbers have on the plasma. Section 3 shows the effect that various parameters have on the ELM frequency. Section 4 presents the results of modelling that takes into account the plasmas response. Section 5 looks at the effect the RMPs have on other plasma parameters and section 6 presents a summary and discussion of the results.

## 2. Characteristics of the single null divertor H-mode plasmas and the RMP configurations used

The baseline plasma scenario for the results presented in this paper is a Lower SND (LSND) plasma, with a plasma current ($I_P$) of 600 kA, a toroidal magnetic field ($B_T$) of 0.55 T at a radius of 0.8 m, a line average density of $4x10^{19}$ m$^{-3}$, and heated by 3.6 MW of Neutral Beam Injected (NBI) power. The plasma enters into a type I ELMing regime at a time of 0.24 s. During the H-mode period the edge safety factor ($q_{95}$) is 2.8, the stored energy is ~ 50-60 kJ, with a normalised plasma pressure ($\beta_N$) of 3.4-4.0.

### 2.1 Effect of RMPs on plasma rotation

Figure 2 shows an example of the effect the application of the RMPs with different toroidal mode numbers (n=3, 4 and 6) has on this LSND H-mode plasma. Figure 2c shows the target $D_\alpha$ intensity as a function of time for the shot with no applied RMPs, where the natural type I ELM frequency ($f_{ELM}$) is ~ 60 Hz.  The radial profiles of the toroidal velocity measured using charge exchange recombination spectroscopy (CXRS) at 10 ms time intervals are shown in Figure 2g.  The profiles show a gradual decrease in the rotation across the whole profile during the time period shown with the core rotation decreasing from 70 to 45 kms$^{-1}$.  The reduction in rotation stops at ~ 0.34 s and from this time onwards the core rotation maintains a constant value of ~ 40-45 kms$^{-1}$ until the plasma current is ramped down at t=0.45-0.5s.  This change in rotation is not due to a mode in the plasma but could be due to an interaction of the plasma with the non-axisymmetric field due to the intrinsic error fields and error field correction coils.

Figure 2d shows the $D_\alpha$ trace for the shot where 1.2 kA have been applied to the 4 turns of the ELM coils giving an effective current of 4.8 kAt in a n=3 configuration.  There is a brief increase in ELM frequency, together with a decrease in the plasma density but this is then followed by a back transition to L-mode.  In this n=3 configuration, the RMPs cause a large braking of the toroidal plasma rotation, which is observed to extend all the way into the core of the plasma (Figure 2h), where within 30 ms of the RMPs being applied the core rotation has been reduced to zero.

Figure 2e shows the $D_\alpha$ trace for the case where the RMPs are applied with 5.6 kAt in an n=4 configuration.  Soon after the coils reach their flat top value there is an increase in $f_{ELM}$ to 230 Hz and consequent decrease in ELM size ($\Delta W_{ELM}$), consistent with $f_{ELM}.\Delta W_{ELM}$~constant.  At the end of the time period shown the line average density has decreased by ~25 % with respect to the shot with no RMPs.  This decrease in line average density is not inconsistent with what be expected from the increase in ELM frequency. As will be discussed in section 5.1 there is also a reduction in the peak heat flux to the target. The rotation profiles shown in Figure 2i again show substantial core braking but the core rotation never reaches zero and instead attains a plateau level of ~ 5kms$^{-1}$ which is sufficient to avoid a back transition to L-mode.

Figure 2f shows the $D_\alpha$ trace for the case where the RMPs are applied with 5.6 kAt in an n=6 configuration. Similar to the n=4 case the application of the RMPs produces an increase in $f_{ELM}$ to ~ 200 Hz and consequent decrease in ELM size and line average density. The rotation profiles shown in Figure 2j again show some core breaking but it is much less than in the n=3 or n=4 cases and the core rotation decreases to a saturated level of ~20 kms$^{-1}$. Repeat shots performed with different currents in the ELM coils showed that the minimum core velocity decreases as $I_{ELM}$ is increased above a threshold value. Due to the lower relative intensity of the charge exchange light to the background light, the CXRS system can not resolve the velocity in the pedestal region as well as it can in the core. However, the measurements that exist suggest that the toroidal rotation at the top of the pedestal remains effectively unchanged at 5 kms$^{-1}$ throughout the time period and is irrespective of the RMP applied. It was not possible to find a coil current for the n=3 configuration that had an effect on the ELM frequency but did not end up producing a back transition to L-mode, so in the rest of the paper only the n=4 and n=6 configurations are studied.

The ERGOS code (vacuum magnetic modelling) [10] has been used to calculate the magnetic perturbations to the plasma due to the coils. The method of using ERGOS on MAST plasmas has been previously described in reference [9]. In all cases only the dominant harmonic is considered (i.e. n=3, 4 or 6) and other harmonics are neglected. Figure 3 a, b and c show the poloidal magnetic spectra of the normalised component of the perturbed field perpendicular to equilibrium flux surfaces ($b^1$) [11] as a function of poloidal mode number (m) and normalised radius ($\Psi_{pol}^{\frac{1}{2}}$) for the n=3, 4 and 6 RMP perturbations respectively. Superimposed on the spectra are the locations of the q=m/n rational surfaces.

The peaks in the applied perturbation are well aligned with the location of the rational surfaces near the edge of the plasma in all three configurations. In the core of the plasma the peak of the perturbations moves further away from the rational surfaces as the toroidal mode number increases from n=3 to n=4 to n=6 hence giving less resonant field components nearer to the core for the higher n RMPs.

The radial profile of effective radial resonant field component ($b^r_{res}$) (see page 47 of reference [11]) for the different RMPs configurations is shown in Figure 4a. The n=6 RMP configuration gives the largest resonant field component at the plasma edge and it also falls off more quickly moving towards the core of the plasma. The n=4 configuration gives a lower value of $b^r_{res}$ at the edge of the plasma but falls off less steeply than the n=6 one and crosses the n=6 curve at $\Psi_{pol}^{\frac{1}{2}} = 0.79$. The n=3 configuration gives the lowest value at the plasma edge but is greater than the n=6 for $\Psi_{pol}^{\frac{1}{2}} < 0.7$ and greater than the n=4 for $\Psi_{pol}^{\frac{1}{2}} < 0.62$. Hence the n=6 configuration is clearly the best for optimising the perturbation at the plasma edge whilst minimising the perturbation in the core, consistent with the effects observed on the plasma rotation (i.e. larger core braking in n=3).

Resonant surfaces are characterized by q=m/n and the Chirikov parameter ($\sigma_{Chirikov}$), which is a measure of the island overlap, is calculated in-between each pair of resonant surfaces as: $\sigma_{Chirikov} = (\delta_m + \delta_{m+1})/\Delta_{m,m+1}$, where $\delta_m$ and $\delta_{m+1}$ represent the half-widths of the magnetic islands on the q=m/n and q= (m+1)/n surfaces (m being the poloidal mode number and q the safety factor) and $\Delta_{m,m+1}$ the distance between these two surfaces. The Chirikov parameter profiles for all 3 configurations are shown in Figure 4b. For higher n the m/n rational surfaces are closer together which naturally results in a larger value of the Chirikov parameter for the higher n RMPs. The region for which the Chirikov parameter is greater than 1 ($\Delta_{\sigma Chirikov>1}$) is used to define the stochastic layer [10], which for the n=3 (4) {6} RMP configurations is 0.067 (0.10) {0.145} in units of $\sqrt{\psi_{pol}}$ or 0.13 (0.19) {0.27} in units of $\psi_{pol}$ respectively.

## 2.2 Effect of RMPs on the L-H transition

Previous studies on MAST [8] have shown that if RMPs in an n=1, 2 or 3 configuration are applied before the L-H transition, with sufficient strength, they can suppress the L-H transition. In order to re-establish the H-mode at the same time the heating power had to be increased by ~ 80 %, if a delay in the L-H transition can be tolerated then the input power had to be increased by ~ 30 %. In contrast, it has been found that the n=4 and n=6 configurations have little effect on the L-H transition but still manage to mitigate the first ELM.

Figure 5 shows a set of shots using RMPs with different toroidal mode numbers. In each configuration the perturbation was applied such that the perturbation had reached flat top well before the time of the L-H transition in the shot without RMPs (Figure 5b). The current in the coils was increased in steps of 1 kAt from one shot to the next until the perturbation was sufficient to suppress the L-H transition at a set input power ($P_{NBI}$ =3.2 MW) or in the case of the n=4 and n=6 configurations the maximum coil current was achieved. The current required to suppress the L-H transition was 4 kAt for the n=3 configuration (Figure 5c). In the case of the n=4 and n=6 configurations the maximum current of 5.6 kAt was insufficient to suppress the L-H transition. Figure 5d and e show the $D_\alpha$ traces for cases where the n=4 and n=6 configurations of the RMPs have been applied before the L-H transition. The L-H transition time is similar to the shot without RMPs (Figure 5b) and the first ELM is mitigated. As can be seen from Figure 5d and e, just after the L-H transition there are some dithery H-mode periods. To investigate whether this was due to proximity to a threshold effect (since the ELM coil current could not be increased) the input power was decreased to 2.4 MW. In this case the result was similar i.e. the L-H transition was still established and the same dithery period existed. Hence it would seem that this is a feature of the early H-mode periods with RMPs applied rather than evidence that the RMPs in the n=4 or n=6 configurations are increasing the power required to achieve H-mode.

A similar suppression of the L-H transition can be produced using an n=2 perturbation, with sufficient strength, from the external error field correction coils. A parameter has not been identified from the vacuum modelling, which can explain why the

n=2 and n=3 configurations suppress the L-H transition whilst the n=4 and n=6 do not. For example, the value of $b^r_{res}$ at the edge of the plasma required to suppress the L-H transition is $0.3x10^{-3}$ for the n=2 configuration and $0.7x10^{-3}$ for the n=3 configuration. While in the n=4 and n=6 configurations, where the L-H transition is not suppressed, the edge value of $b^r_{res}$ is $1.4x10^{-3}$ and $1.8x10^{-3}$ respectively.

Hence, at least from vacuum modelling no clear marker can be found to identify when the L-H transition will be suppressed. On the positive side, empirically it does appear to be more difficult to suppress the L-H transition at higher n, which may be good news for machines like ITER which will operate in an n=4 configuration.

## 3. Effect of RMPs on ELM frequency

### 3.1 Effect of ELM coil current

Repeat discharges have been performed with increasing current in the coils ($I_{ELM}$) to determine the threshold current for the onset of ELM mitigation together with the effect on ELM frequency. Figure 6 shows a series of shots with increasing $I_{ELM}$ for the RMPs in an n=4 configuration. Figure 6c shows the $D_\alpha$ trace for the shot with $I_{ELM}= 0$, which has $f_{ELM}$ =55Hz. The ELM coil current has been increased in subsequent shots in steps of 0.4 kAt. For $I_{ELM}< 2.4$ kAt no effect on the ELM frequency is observed. Figure 6d shows the $D_\alpha$ trace for a shot with $I_{ELM}=4.0$ kAt, the ELM frequency, averaged over the time the coils current is in flat top, has increased to 130 Hz. Figure 6e and f show the $D_\alpha$ traces for shots with $I_{ELM} = 4.8$ and 5.6 kAt respectively where $f_{ELM}$ continues to increase with $I_{ELM}$. At the maximum ELM coil current (5.6 kAt) the ELM frequency is 290 Hz which is a factor of 5.3 times larger than the natural $f_{ELM}$.

The same scan has been performed with the RMPs in an n=6 configuration. In this configuration the threshold current required to increase the ELM frequency is $I_{ELM}$= 3.6 kAt, which is higher than that found for the n=4 configuration. Figure 7a summarises $f_{ELM}$ as a function of $I_{ELM}$ for the n=4 and n=6 configurations. While the different threshold current in the two cases is visible, the rate of increase of $f_{ELM}$ with $I_{ELM}$ above the respective thresholds is similar. In order to see if an ordering parameter can be found based on vacuum modelling, ERGOS simulations have been performed for these shots using the respective coil configurations. Two parameters chosen are the maximum value of $b^r_{res}$ at the edge (Figure 7b) and the region over which the Chirikov parameter is greater than 1 ($\Delta_{\sigma Chirikov>1}$) in units of $\psi_{pol}$ (Figure 7c). However, there does not seem to be a single parameter since the threshold value for the onset of ELM mitigation is $b^r_{res} \sim 0.4 \times 10^{-3}$ for n=4 and $0.7 \times 10^{-3}$ for n=6 and $\Delta_{\sigma Chirikov>1} \sim 0.12$ for n=4 and 0.18 for n=6.

### 3.2 Effect of density/refuelling

While clear ELM mitigation has been observed, ELM suppression has not been established. Since ELM suppression has been established in DIII-D at high and low collisionality [2][4], while only ELM mitigation is observed at intermediate values, and keeping in mind that there is a density threshold for complete suppression of type I ELMs on ASDEX Upgrade [3], a scan in fuelling rate and density has been performed.

Figure 8 shows pairs of shots with $I_{ELM} = 0$ or 4 kAt for the RMPs in an n=4 configuration. Figure 8 c,d (e,f) {g,h} show the $D_\alpha$ traces for the pairs of RMP off/on shots for the three different refuelling rates used. In the case of no refuelling (Figure 8 c,d) the type I ELM frequency increases from the natural value of $f_{ELM} = 60$ Hz for $I_{ELM} = 0$ to $f_{ELM} = 165$ Hz for $I_{ELM}$=4 kAt.

As on most devices the power required to achieve H-mode is minimum at a certain density (non-zero); below this density on MAST the L-H transition power increases rapidly. As can be seen from Figure 8 b in the RMP applied case there is a considerable decrease in the density, this decrease eventually leads to a back transition to L-mode as the minimum in the $P_{L-H}$ versus density point is reached. In order to compensate this density pump out, and hence avoid this back transition, different refuelling rates have been used. Figure 8e and f show the pair of shots where a moderate refuelling rate of $4.6\text{x}10^{21}$ $D_2$ $s^{-1}$ has been applied from 0.3 s. The natural ELM frequency only increases marginally to $f_{ELM} = 80$ Hz and the mitigated frequency increases to 195 Hz and there is little effect on the pedestal temperature. The density decrease in the RMP on shot has now been compensated and the increase in density in the RMP off shot is small. Increasing the gas refuelling further to a higher rate of $8\text{x}10^{21}$ $D_2$ $s^{-1}$ produces a degradation in the confinement of the shot, the temperature pedestal decreases and although $f_{ELM}$ increases to 290 Hz in the RMP applied shot it also increases in the RMP off shot ($f_{ELM}$ =145 Hz). A summary of $f_{ELM}$ versus fuelling rate is shown in Figure 9. The mid refuelling rate of $4.6\text{x}10^{21}$ $D_2$ $s^{-1}$ was found to be the best compromise for reducing the density pump out and hence avoiding the back transition whilst still retaining the good confinement of the underlying plasma.

The electron collisionality at the top of the pedestal has been calculated following reference [12] as:

$$\nu_e^* = 6.921.10^{-18} \frac{Rqn_e Z_{eff} \ln \Lambda_e}{\varepsilon^{3/2} T_e^2}$$

where R is the major radius in m, q is the safety factor at the pedestal top, $\varepsilon$ is the inverse aspect ratio, $Z_{eff}$ is the effective ion charge, $n_e$ the electron density in $m^{-3}$ and $T_e$ the temperature in eV, both evaluated at the top of the pedestal. $\ln \Lambda_e$ is the Coulomb logarithm defined by $\ln \Lambda_e = 31.3 - \ln(\sqrt{n_e}/T_e)$. In the RMP off shots shown in Figure 8 the pedestal top collisionality increases from 0.5 in the non fuelled case (Figure 8c), to 1.0 in the mid refuelling case (Figure 8e) up to a maximum value of 1.8 in the high refuelling case (Figure 8g). The pedestal top density increases from $4.0 \times 10^{19}\,m^{-3}$ to $4.5 \times 10^{19} m^{-3}$ with the majority of the increase in collisionality being due to a decrease in the pedestal temperature.

Shots have also been performed at different initial densities. Figure 10a shows that a wide range in pedestal top collisionality ($\nu_e^*$) has been explored from $0.4 < \nu_e^* < 2.0$. The lower limit is set by the minimum density required to achieve H-mode at the available heating power while the upper limit is set by the maximum density that can be achieved whilst maintaining the plasma in a type I ELM-ing regime. Unfortunately, the collisionality range scanned coincides with the window for which ELM suppression is not observed in DIII-D. On ASDEX Upgrade the suppression of type I ELMs is not associated with collisionality (although the collisionalities in the MAST discharges overlap those in ASDEX Upgrade 0.8-2.0 ), but rather the plasma density expressed as a fraction of the Greenwald density ($n_{GW}$), with suppression being observed for $ne/n_{GW}>0.53$ [13]. Figure 10b shows the distribution of the Greenwald density fraction for the MAST discharges. While most lie in the range 0.2 $< n_e/n_{GW} < 0.4$ a few discharges have been performed in the range that overlap with the ASDEX Upgrade type I ELM suppression region. However, on MAST no suppression is observed, although the ELM frequency increases.

### 3.3 Effect of $q_{95}$

In order to test the sensitivity of the ELM frequency to alignment of the applied perturbation with the pitch of the equilibrium magnetic field (i.e. to test if a resonant condition exists) a scan in $q_{95}$ has been performed by repeating discharges with different values of the toroidal field. Figure 11a shows the results from repeat discharges with $B_T$ in the range 0.48 to 0.585 T, corresponding to $q_{95}$ in the range 2.4 to 3.0. For this range of $B_T$ there is very little change in natural ELM frequency, although both the n=4 (performed with $I_{ELM}$ = 4.0 kAt) and n=6 (performed with $I_{ELM}$=5.6 kAt) perturbations show a dependence of $f_{ELM}$ on $q_{95}$  The largest ELM frequency is obtained with $q_{95}$ = 2.6 ($B_T$=0.52 T) for the n=4 configuration and $q_{95}$ = 2.8 ($B_T$=0.55 T) for the n=6 configuration. There is a clear dependence of $f_{ELM}$ on $q_{95}$, suggesting that the ELM frequency is sensitive to the alignment of the applied perturbation with the pitch of the equilibrium magnetic field. RMP experiments performed on JET also observed a dependence of $f_{ELM}$ on $q_{95}$ [14], however, in that case a multi-resonant effect is observed where the increase in ELM frequency is cyclical. The difference in $q_{95}$ between two neighbouring peaks in ELM frequency is in a range of $\Delta q_{95} \sim 0.2$–$0.3$ [14]. There is no evidence for such a multi-resonant effect in the limited q95 range studied in the current MAST data.

If the dependence of $f_{ELM}$ on $q_{95}$ observed in the MAST data was only related to the alignment of the perturbation with the equilibrium field then it may be possible to identify a parameter from modelling that would quantify this alignment. A search for such a parameter has been performed using vacuum modelling and the maximum values of $b^r_{res}$ has been calculated as a function of q95 for the n=4 and n=6 configurations of the coils. The lack of correlation observed in a plot of $f_{ELM}$ versus $b^r_{res}$ (Figure 11b) could be due to the fact that the vacuum approximation is not appropriate and/or because even though all the non-axisymmetric fields are included in ERGOS only the dominant components (i.e. n = 4 or n=6) are used when deriving these variables or as suggested in reference [14] there are other effects which determine the dependence of $f_{ELM}$ on $q_{95}$.

### 3.4 Effect of distance between the plasma and the RMP coils

Another way of varying the size of the applied RMP is to vary the distance of the plasma from the coils. Figure 12 shows a series of shots with the coils in an n=6 configuration with different distance to the coils. Figure 12d and e show the $D_\alpha$ traces for a pair of shots with $I_{ELM} = 0$ and 5.6 kAt where the outer radius of the plasma at the mid-plane is 1.4 m (Figure 12c) corresponding to the poloidal cross section shown as a solid line in Figure 1. Figure 12f shows the $D_\alpha$ trace for a shot again with $I_{ELM} = 5.6$kAt but where the plasma outer radius has been reduced to 1.3 m, corresponding to the poloidal cross section given by the dashed curve in Figure 1. As can be seen in the reduced outer radius case the RMPs have no effect on the ELM frequency.  However, they do still cause substantial braking of the toroidal rotation of the plasma, with the core rotation decreasing to 25 kms$^{-1}$, similar to that observed in the larger radius shots.  This shows that the change in ELM frequency is not simply due to the change in rotation.

Figure 12g shows the D$_\alpha$ trace for a shot that starts at the larger radius and during the application of the RMPs the plasma radius is reduced (Figure 12c).  In spite of the fact that the ELM mitigated stage has been established as the gap to the coils is increased the ELM frequency decreases until the natural value of f$_{ELM}$ is established.

Figure 13a shows f$_{ELM}$ for a series of shots performed at different outer radii.  There is clearly a threshold value of the outer radius below which there is no effect of the RMPs on f$_{ELM}$.  Vacuum modelling has been used to calculate the value of b$^r_{res}$ for these different discharges.  Figure 13b shows that f$_{ELM}$ increases linearly above a threshold value of b$^r_{res}$ = $1.1 \times 10^{-3}$.  This threshold value is different to that ($0.7 \times 10^{-3}$) found for the n=6 configuration during the I$_{ELM}$ scan performed at fixed radius shown in Figure 7b.  Again, this indicates that there is not a single parameter from vacuum modelling which can be used to determine the threshold for the onset of ELM mitigation.

## 4.  Plasma response modelling

Calculations have been performed using the MARS-F code, which is a linear single fluid resistive MHD code that combines the plasma response with the vacuum perturbations, including screening effects due to toroidal rotation [15].  The calculations use the experimental profiles of density, temperature and toroidal rotation as input and realistic values of resistivity, characterised by the Lundquist number (S) which varies from $\sim 10^8$ in the core to $\sim 10^6$ in the pedestal region (the radial profile of the resistivity is assumed proportional to $T_e^{-3/2}$). The resistive plasma response significantly reduces the field amplitude near rational surfaces and reduces the resonant component of the field by more than an order of magnitude (Figure 14a) resulting in similar values for both the n=4 and n=6 RMP configurations. In the MARS-F modelling the RMP field also causes a 3D distortion of the plasma surface (Figure 14b), which potentially leads to the formation of a 3D steady state equilibrium. The plasma displacement varies with toroidal angle ($\phi$) as $\xi e^{in\phi}$ where n is the toroidal mode number and $\xi$ is the amplitude of the normal displacement of the plasma surface which varies as a function of poloidal angle ($\theta$).

Previous MARS-F simulations of the effect of RMPs on the MAST plasma showed a clear correlation between the location of the maximum of the amplitude of the normal component of the plasma displacement at the plasma surface and the effect of the RMPs on the plasma [16]. In these studies it was observed that a density pump out in L-mode or ELM mitigation in H-mode only occurred when the displacement at the X-point was larger than the displacement at the mid-plane.

Figure 14b shows this displacement ($\xi$) as a function of $\theta$ for the n=4 and n=6 coil configurations calculated at $\Psi_{pol} = 0.98$. In both cases the displacement has a peak near the X-points ($\theta \sim -90^\circ$), however, for the same $I_{ELM}$ the displacement is larger in the case of n=4 than n=6. The maximum displacement is of the order of 0.7 mm/kAt i.e. for the coil current used (5.6kAt) the maximum displacement is $\sim$ 4mm.

The MARS-F calculations have been performed for the discharges used in the $I_{ELM}$ scan shown in Figure 7a and the value of $b^r_{res}$ taking into account the plasma response and the displacement ($\xi$) have been calculated. Figure 15a shows $f_{ELM}$ versus $b^r_{res}$ with the plasma response included. Similar to what was observed in the vacuum approximation calculations, there is a different threshold in the n=4 and n=6 cases. Figure 15b shows a plot of $f_{ELM}$ versus the displacement at the X-point, this results in a collapse of the data into a single trend for both the n=4 and n=6 configurations, with a single threshold of ~1.5 mm for the onset of ELM mitigation.

The quasi-linear MARS-Q code [17] has been used to simulate the RMP penetration dynamics and the toroidal rotation braking for the shots shown in Figure 2. The MARS-Q code employs a full MHD, single fluid model for the plasma response in a full toroidal geometry. The MHD equations are solved in the time domain, including both the fluid $\vec{j} \times \vec{B}$ and the NTV torque in the momentum balance equation. The CXRS system, used to produce the rotation profiles shown in Figure 2, measures at the mid-plane of the vessel (Z=0), therefore, for these LSND shots it does not measure at the magnetic axis. The innermost point corresponds to a normalised poloidal flux ($\Psi_{pol}$) of 0.2-0.3. Figure 16a shows the time evolution of the toroidal velocity measured at $\Psi_{pol}$ =0.3 for the three RMP configurations shown in Figure 2 as a function of time after the RMPs have reached flat top. Since the MARS-Q modelling does not know about changes in torque other than that due to the applied RMPs, the toroidal rotation velocities have been corrected for the gradual decrease in the toroidal velocity that is observed in the shot without RMPs. In all the shots with RMPs applied there is considerable braking of the core rotation. In each case the deceleration of the plasma is similar, with just the saturated level being different for the three cases i.e. $V_\phi^{min}$ = 0 (n=3), 5 (n=4) and 20 (n=6) kms$^{-1}$.

Figure 16b shows the MARS-Q simulated toroidal rotation at $\Psi_{pol} = 0.3$ as a function of time after the RMPs have been applied. The assumptions made in these simulations are given in reference [18]. For the n=3 configuration of the RMPs the code predicts a full damping of the toroidal rotation, mainly due to the $\vec{j} \times \vec{B}$ torque, in a time of less than 40ms, very similar to what is observed in the experiment (Figure 16a). For the n=4 and n=6 configurations a similar rate of damping is predicted together with a prediction that a minimum saturated level would be achieved. The value of the saturated level is in good agreement for the n=6 configuration but the code predicts a higher saturated level for the n=4 configuration that what is observed experimentally. This could be due to the fact that the simulation only includes the n=4 component of the applied field whereas, as was discussed in section 1, the n=4 coil configuration also has a sizeable n=8 sideband.

## 5. Effect of RMPs on other plasma parameters

### 5.1 Effect on ELM size and target heat loads

Figure 17a shows a plot of the energy loss per ELM ($\Delta W_{ELM}$), derived from the change in plasma stored energy calculated by the EFIT equilibrium code [19], versus $f_{ELM}$ for the natural and mitigated ELMs. The application of the RMPs produces an increase in $f_{ELM}$ and corresponding decrease in $\Delta W_{ELM}$ consistent with $f_{ELM}.\Delta W_{ELM} =$ const (represented by the solid curve in Figure 17a). The change in plasma stored energy (averaged over the ELM cycle) decreases by between 5 to 10 kJ in the ELM mitigated shots compared to the shots without RMPs applied, which represents a decrease in confinement of between 8 and 16 %.

The ELM energy loss is often discussed in terms of convective (characterised by changes in the plasma density) and conductive losses (characterised by changes in the plasma temperature) such that the ELM energy loss expressed as a fraction of the pedestal energy ($W_{ped}$) can be written as

$$\frac{\Delta W_{ELM}}{W_{ped}} = \frac{\Delta n_e}{n_e^{ped}} + \frac{\Delta T}{T^{ped}}$$

with the smallest ELM sizes being observed when $\Delta T = 0$ (see [20] and references therein). The convected ELM energy loss is typically ~ 3-4% in all devices and usually remains constant as the density and collisionality are varied in any given machine and configuration [20]. If this held true for mitigated ELMs it would place an ultimate limit on the smallest ELM size achievable irrespective of the ELM frequency achieved. Fortunately this is not the case as can be seen in Figure 17b, which shows $\Delta n_e/n_e^{ped}$ as a function of $f_{ELM}$. The natural ELMs have a mean value of $\Delta n_e/n_e^{ped} = 0.04$ ,whereas for the mitigated ELMs $\Delta n_e/n_e^{ped}$ decreases with increasing ELM frequency.

In order to avoid damage to in-vessel components in future devices, such as ITER, it is the peak heat flux density at the divertor that is important rather than the actual ELM size. The divertor heat fluxes on MAST have been measured using infrared thermography. Figure 17c shows the peak heat flux density at the target ($q_{peak}$) as a function of $\Delta W_{ELM}$. The increase in ELM frequency and decrease in $\Delta W_{ELM}$ does lead to reduced heat fluxes at the target, although it also results in a smaller wetted area at the target meaning that the reduction in $q_{peak}$ is not the same as the reduction in $\Delta W_{ELM}$. The mitigated and natural ELMs follow the same trend and show that a reduction of a factor of 5 in $\Delta W_{ELM}$ (i.e. from 15 to 3kJ) produces a reduction in $q_{peak}$ of 1.8 (from 22 to 12 MWm$^{-2}$). In order to make extrapolations over a wider range it will be important to understand what happens at very small energies, since a linear extrapolation based on the present data would indicate a non zero heat flux for $\Delta W_{ELM} = 0$.

## 5.2 Effect on ELM filaments

Filament structures have been observed during ELMs in a wide range of Tokamaks using a variety of diagnostics (see [21] and references therein). Results from coordinated experiments on ASDEX Upgrade and MAST have shown that the toroidal mode number derived from the analysis of these images can be a good indicator of the ELM type [22]. Measurements have been performed on MAST using images obtained from a Photron Ultima APX-RS camera, which was used to continuously record unfiltered light, dominated by $D_\alpha$ emission, throughout the entire shot. In this present analysis it has been used with a 5 µs exposure time in two modes: either a full plasma view (512x462 pixels) at 7.5 kHz framing rate or a view of a region around the Low Field Side (LFS) mid-plane region of the plasma (256x48 pixels) at 100 kHz.

Figure 18a and b show a typical full frame images obtained using a 5 µs exposure during the rise time of the mid-plane $D_\alpha$ signal for a natural ELM ($I_{ELM} = 0$) and a mitigated ELM obtained in a discharge with the RMPs in an n=6 configuration with $I_{ELM}$=5.6 kAt. The ELM energy loss ($\Delta W_{ELM}$) in the two cases is 8.0 and 2.0 kJ respectively. The raw image obtained for the mitigated ELM is dimmer due to the smaller

ELM loss. The image shown in Figure 18b has been brightened to give the same intensity as Figure 18a to enable an easier comparison of the filamentary structures, which are qualitatively similar for both ELMs.

The high frame rate mid-plane view has been used to measure the mode number and propagation of the filaments during each ELM by determining the toroidal and radial location of each filament in subsequent frames, separated by 10 μs. The images have been analysed by mapping 3-D field lines, generated from the magnetic equilibrium, onto the 2-D image [23] and the intensity along the field line calculated as a function of toroidal angle. A peak finding detection algorithm is then applied to the trace of intensity versus toroidal angle and results in the toroidal location and the half width half maximum (HWHM) toroidal extent of the filaments being determined [22]. The same technique is then applied to subsequent frames to determine the toroidal propagation of each filament.

Measurements of the separation and toroidal propagation of the filaments while they remain at the LCFS have been repeated for all the ELMs in a series of shots with and without RMPs. Figure 19a shows the probability distribution function for the toroidal velocity ($V_\phi$). In all cases the filaments start off rotating at a constant toroidal velocity but decelerate toroidally before they move radially outwards. Therefore the toroidal velocity plotted in Figure 19a is that obtained during the initial stage when the toroidal velocity is constant. Although there is considerable core rotation braking when the RMPs are applied the velocity distribution of the filaments is similar with and without RMPs. The measured toroidal velocity of ~ 5 kms$^{-1}$ is similar to the toroidal rotation velocity of the pedestal in these discharges, which is measured using charge exchange recombination spectroscopy to be ~5 kms$^{-1}$.

The mean separation in the toroidal angle between the filament locations is used to derive the toroidal mode number (n), which is shown in Figure 19b. In all cases the toroidal mode number derived is similar with a mean value of 13, similar to that found for type I ELMs in other discharges in MAST [22]. Finally the toroidal width of the filaments has been determined from the width of the intensity distribution. Figure 19c shows that the filament widths are not affected by the application of the RMPs.

The analysis of the filament images suggests that, similar to what was observed in the CXRS data; the application of the RMPs does not modify the edge toroidal rotation. The toroidal mode number of the ELMs is also unaffected and the mode number for the natural and mitigated ELMs is consistent with what is expected from type I ELMs i.e. the increased ELM frequency in the mitigated stage is not due to a transition to type III ELMs which in MAST have been shown previously to have a higher mean mode number (n=20) and wider distribution of mode numbers (from n=5 to 30) [22].

### 5.3 Effect on pedestal parameters

The pedestal electron density and temperature characteristics have been measured using a Nd YAG Thomson Scattering (TS) system. The radial pedestal profiles for shots without RMPs and with RMPs in an n=4 and n=6 configuration, obtained in the last 10 % of the ELM cycle, are shown in Figure 20a, b and c for the electron density, temperature and pressure respectively mapped onto normalised flux, using the unperturbed equilibrium. In radial space a shift of the pedestal profile is observed of the order of ~1cm due to the application of the RMPs [24]. To compensate for this displacement when mapping to poloidal flux, the profiles have been aligned using a constraint based on the power crossing the Last Closed Flux Surface (LCFS) that sets the electron temperature at the LCFS to be ~40 eV. A clear drop is observed in the pedestal density but little change in the electron temperature. The pedestal top pressure reduces and the pedestal width increases resulting in a decrease in the peak pressure gradient. The broadening of the pressure pedestal width is different to the behaviour observed on DIII-D [5] and ASDEX Upgrade [13] where no change in the pedestal width is observed.

A stability analysis has been performed on these discharges using the ELITE stability code [25]. The procedure used to analyse the edge stability in MAST has been described in [26]. It consists of reconstructing the equilibrium using the kinetic profiles obtained from the TS system as constraints and assuming that $T_i = T_e$. The current profile is calculated by combining the bootstrap current, calculated using the formula given by Sauter [12], and the ohmic current. The edge pressure gradient is then varied at a fixed current density and the edge stability evaluated using ELITE [25].

Figure 21 shows the stability boundary and the experimental point in a plot of peak edge current density ($j_\phi$) versus normalised pressure gradient ($\alpha$) for the discharge without RMPs and for the discharge with the RMPs in an n=6 configuration (the n=4 configuration gives a similar result). The results show that for the discharge without RMPs the experimental point lies in the region unstable to peeling-ballooning modes, a trait often associated with type I ELMs. However, for the point with RMPs the analysis predicts that such a discharge would be stable to peeling-ballooning modes and so it is not apparent why the ELM frequency should be higher in such a discharge. However, such a stability analysis assumes toroidally symmetric and smooth edge flux surfaces, but as will be discussed below, during the application of RMPs the edge is anything but smooth and maybe it is these deformations of the surface that lead to greater instability.

## 5.4 Effect on the X-point

The lower X-point region of the plasma has been viewed using a toroidally viewing camera with a spatial resolution of 1.8mm at the tangency plane. The image has been filtered with either a He II (468 nm) or a CIII (465 nm) filter and the images obtained using an integration time of 3000 or 300 μs respectively. These lines have been chosen since they are the strongest impurity lines in the typical plasma conditions found at the plasma boundary. Figure 22a shows a false colour image obtained using a He II filter at 0.32s in the shot with $I_{ELM}$ =0 kAt during an inter-ELM period. The image shows a smooth boundary layer associated with the LCFS. In contrast, Figure 22b and c show an images obtained at the same time during an inter-ELM period for a shot with $I_{ELM}$=5.6 kAt with the coils in an n=6 and n=4 configurations respectively. Clear lobe structures are seen near to the X-point. The location and poloidal separation of the lobes is different for each toroidal mode number of the perturbation.

In an ideal axi-symmetric poloidally diverted tokamak the magnetic separatrix (or LCFS) separates the region of confined and open field lines. Non-axi-symmetric magnetic perturbations split this magnetic separatrix into a pair of so called "stable and unstable manifolds" [27]. Structures are formed where the manifolds intersect and these are particularly complex near to the X-point. The manifolds form lobes that are stretched radially both outwards and inwards. Some of these lobes can intersect the divertor target and result in the strike point splitting often observed during RMP experiments [28][29]. In reference [30] it is shown that the radial extent of the lobes sets a minimum value on the radial extent of the stochastic layer, i.e. the stochastic layer has to be at least as broad as the lobes.

A good quantitative agreement has been reported [31] between the number and separation of the lobes in the image and the vacuum modelling performed using the ERGOS code [10]. However there appears to be a discrepancy in their radial extent. The size of the lobes has been determined by mapping onto the image contours at fixed radial distances from the unperturbed last closed flux surface. The radial extent of the longest finite length lobe (i.e. not those that extend down to the divertor) has been measured for repeat shots performed at different values of $I_{ELM}$. Figure 23a shows the lobe length as a function of $I_{ELM}$ for the RMPs in the n=4 and n=6 configurations. For coil currents above a threshold ($I_{THR}$) the extent of the lobes increases approximately linearly with $I_{ELM}$-$I_{THR}$. Lobes are not observed for $I_{THR} \leq 2.4$ kAt in the n=4 and 3.2 kAt in the n=6 configurations respectively. This is similar to the thresholds observed for the onset of ELM mitigation (i.e. the increase in ELM frequency). However, it could be that it is just difficult to measure small lobes and an alternative threshold could be determined by extrapolating a linear fit to the data to zero lobe length. This would give $I_{THR} = 1.2$ kAt for n=4 and 2.0 kAt for n=6.

Figure 23b shows the measured increase in ELM frequency versus lobe length where a linear relationship can be seen. The fact that the size of the lobe length is so well correlated with the change in ELM frequency may suggest that the lobes themselves are having a direct impact on the stability of the edge plasma to peeling ballooning modes. Such 3D perturbations to the separatrix are not included in present stability codes. The stability of the edge plasma has been tested by applying a perturbation to the boundary shape [32], which shows that the presence of such perturbations do indeed degrade the stability to ballooning modes. This degradation in ballooning stability originates from the perturbed field lines dwelling in the region of unfavourable curvature due to the presence of lobe structures rather than the change in the plasma boundary shape. At present these calculations are only a proxy for what is required, which is that the full 3D nature of the perturbations will need to be included in the stability calculations.

The original motivation for viewing the X-point region was to try to visual the MHD displacement predicted by the MARS-F code. The predicted MHD displacement for IELM = 5.6 kAt is $\xi$=4 mm, which is much smaller than the lobe structures which are up to 10cm long. It is not clear how the MHD displacement is related to the lobe structures observed and this will be the subject of further studies.

## 6. Summary and discussion

Experiments have been performed on lower SND MAST plasmas using internal (n=3, 4 or 6) resonant magnetic perturbation coils. Sustained ELM mitigation has been achieved using RMPs with a toroidal mode number of n=4 and n=6. The application of the RMPs produces braking of the toroidal rotation, which in the case of the n=3 configuration is so severe that it produces a back transition to L-mode before any sustained ELM mitigation can be achieved. The ELM frequency increases by up to a factor of five with a similar reduction in ELM energy loss. The peak heat flux at the target also decreases by a factor of 1.8. ELM mitigation has been observed for the whole range of shots covered ($0.4 < v_e^* < 2.0$, $0.2 < n_e/n_{GW} < 0.6$) but ELM suppression has not been observed. The application of the RMPs in the n=4 and n=6 configurations before the L-H transition have little effect on the power required to achieve H-mode while still allowing the first ELM to be mitigated.

Coincident with the effect on the ELMs, clear lobe-like structures are observed near to the X-point. The appearance of these lobes is correlated with the effect of the RMPs on the plasma i.e. they only appear when the RMPs have an effect on the ELM frequency. The structures are not seen if the ELM coil current is too small.

A threshold current for ELM mitigation is observed above which the ELM frequency increases approximately linearly with current in the coils. The threshold current is lower in the n=4 configuration than in the n=6 configuration. Although calculations in the vacuum approximation using the ERGOS code show a linear response above the threshold they can not explain the difference in the threshold. Scans have also been performed in $q_{95}$ and the gap between the plasma and the RMP coils. Both show strong dependencies although the exact dependencies can not be explained by vacuum modelling.

Plasma response calculations performed using the MARS-F code show a strong correlation between the MHD plasma displacement at the X-point and the ELM frequency for n=4 and n=6, with a similar threshold displacement of 1.5 mm in both cases. A similar correlation is also observed between the ELM frequency and the length of the lobes observed near to the X-point.

The single fluid model used in this paper captures the damping of the plasma flow, which in turns enhances the RMP field penetration. It does not capture the subtle physics of 2-fluid effects, which mainly relate to diamagnetic flow effects. In order to fully model the physics of the pedestal region a full nonlinear two-fluid model is most likely required [33][34][35][36]. A future piece of work must involve including 2-fluid and kinetic physics in order to correctly describe the RMP penetration problem and the effect this has on ELM dynamics.

A comparison of the filament structures observed during the ELMs in the natural and mitigated stages show that they both have similar characteristics. Based on the toroidal mode number of the filaments it would appear that the mitigated ELMs still have all the characteristics of type I ELMs even though their frequency is higher, their energy loss is reduced and the pedestal pressure gradient is decreased. The correlation of ELM frequency with the size of the lobes may suggest that these perturbations, which are not included in present stability codes, may be playing a role in destabilising ballooning modes and the inclusion of such effects in future codes may help to explain why the ELM frequency increases.

All the results presented in this paper, suggest that in terms of overall plasma performance it is best to perform ELM mitigation with RMPs with a higher toroidal mode number (i.e. n=4 or 6). This may be good news for machines like ITER that are planned to operate in an n=4 configuration.

**Acknowledgement**

This work was funded partly by the RCUK Energy Programme under grant EP/I501045 and the European Communities under the contract of Association between EURATOM and CCFE. The views and opinions expressed herein do not necessarily reflect those of the European Commission. This work was carried out within the framework of the European Fusion Development Agreement.

**References**

[1] Loarte A *et al.*, 2003 Plasma Phys. Control. Fusion **45** 1594

[2] Evans T E *et al.*, 2004 Phys. Rev. Lett. **92** 235003

[3] Suttrop W *et al.,* 2011 Phys. Rev. Lett. **106** 225004

[4] Evans TE *et al*., 2008 Nucl. Fusion **48** 024002

[5] Snyder PB *et al.* 2012 Phys. Plasmas **19** 056115

[6] Liang Y *et al.*, 2007 Phys. Rev. Lett. **98** 265004

[7] Liang Y *et al.*, 2010 Nucl. Fusion **50** 025013

[8] Kirk A *et al.,* 2011 Plasma Phys. Control. Fusion **53** 065011

[9] Kirk A *et al*., 2010 Nucl. Fusion **50** 034008

[10] Nardon E *et al*., 2007 *J. Nucl. Mater.* **363-365** 1071

[11] Nardon E. 2007 "Edge Localized modes control by resonant magnetic perturbations" *PhD Thesis* Ecole Polytechnique

[12] Sauter O, Angioni C and Lin-Liu Y R 1999 *Phys. Plasmas* **6** 2834

[13] Fischer R et. al., 2012 "Response of kinetic profiles to non-axisymmetric magnetic perturbations at ASDEX Upgrade" 2012 Submitted to Plasma Phys. Control. Fusion

[14] Liang Y *et al.*, 2010 Phys. Rev. Lett. **105** 065001

[15] Liu Yueqiang *et al.*, 2010 *Phys. Plasmas* **17** 122502

[16] Liu Yueqiang *et al.*, 2011 Nucl. Fusion 5**1** 083002

[17] Liu Yueqiang *et al*., 2012 " Toroidal modelling of penetration of resonant magnetic perturbation field" Submitted to Phys. Plasmas

[18] Liu Yueqiang *et al*., 2012 " Toroidal modelling of plasma response and RMP field penetration" Submitted to Plasma Phys. Control. Fusion

[19] Lao LL *et al.,* 1985 Nucl. Fusion **25** 1611

[20] Leonard AW et al., 2006 Plasma Phys. Control. Fusion **48** A149

[21] Kirk A *et al*., 2005 Plasma Phys. Control. Fusion **47** 995

[22] Kirk A *et al*., 2011 Plasma Phys. Control. Fusion **53** 095008

[23] Kirk A *et al*., 2006 Phys. Rev. Lett. **96** 185001.

[24] Chapman IT et al., 2012 Plasma Phys. Control. Fusion **54** 105013

[25] Wilson HR *et al*., 2002 Physics of Plasmas **9** 1277

[26] Saarelma S et al., 2007 Plasma Phys. and Control. Fusion **49** 31.

[27] Wingen A *et al* 2009 Nucl. Fusion **49** 055027

[28] Jakubowski MW *et al*., 2009 Nucl. Fusion **49** 095013

[29] Nardon E *et al.*, 2011 J. Nucl. Materials **415** S914

[30] Abdullaev SS *et al.,* 2002 Physics of Plasmas **15** 042598

[31] Kirk A *et al*., 2012 Phys. Rev. Lett. **108** 255003

[32] Chapman IT et al., 2012 "Towards understanding ELM mitigation: The effect of axisymmetric lobe structures near the X-point on ELM stability" Submitted to Nuclear Fusion

[33] Izzo VA and Joseph I, 2008 Nucl. Fusion **48** 115004

[34] Strauss HR *et al.,* 2009 Nucl. Fusion **49** 055025

[35] Nardon E *et al.,* 2010 Nucl. Fusion **50** 034002

[36] Ferraro NM et al, 2012 Physics of Plasmas **19** 056105

**Figures**

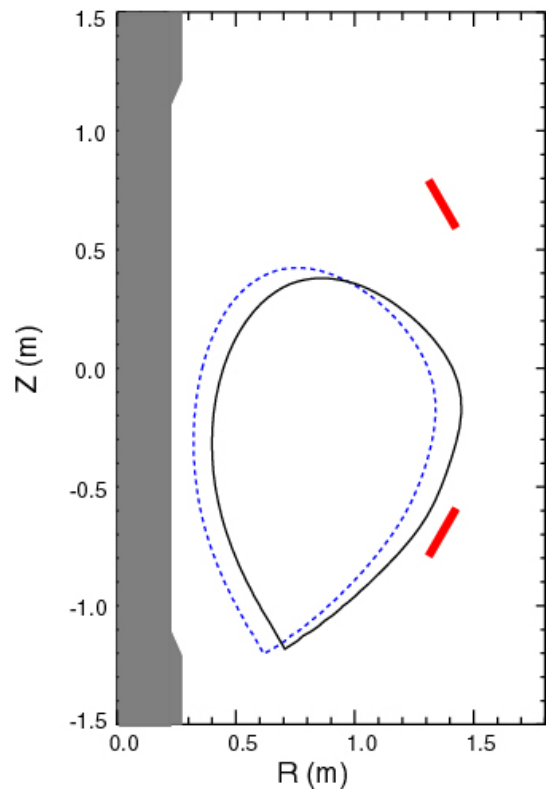

**Figure 1** Poloidal cross section of the standard LSND plasmas used (solid curve) and inward shifted plasma (dashed) together with the location of the centre column and ELM coils.

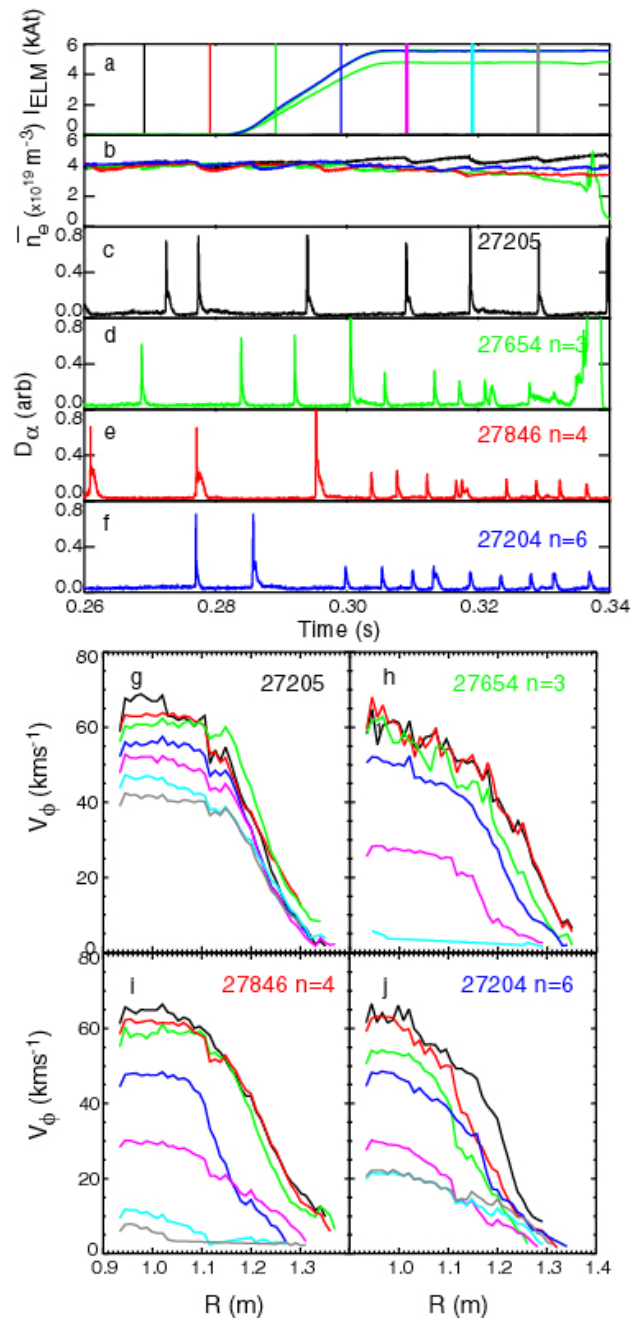

**Figure 2** Time traces of a) the current in the ELM coils ($I_{ELM}$) b) line average density ($\bar{n}_e$), and the target $D_\alpha$ intensity for c) a lower SND shot without RMPs and d),e),f) with RMPs in an n=3,4,6 configuration respectively. The radial profile of the toroidal rotation velocity at 10 ms time intervals in the discharges for g), h), i) and j) shots without RMPs and with RMPs in an n=3, 4 and 6 configurations respectively. The rotation profiles are obtained at the times of the same coloured vertical lines shown in a).

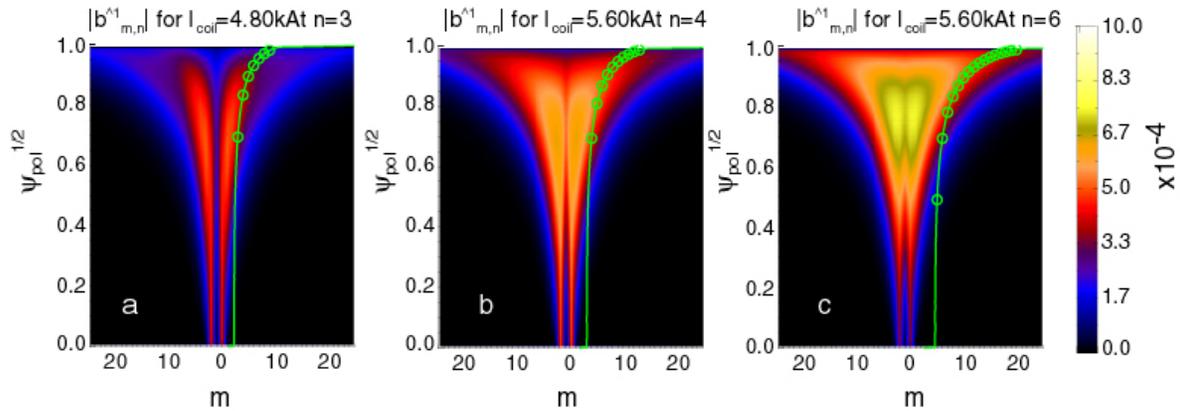

**Figure 3** Poloidal magnetic spectra calculated in the vacuum approximation for RMPs in a) an n=3, b) n=4 and c) an n=6 configuration. Superimposed as circles and solid line are the q=m/n rational surfaces of the discharge equilibrium

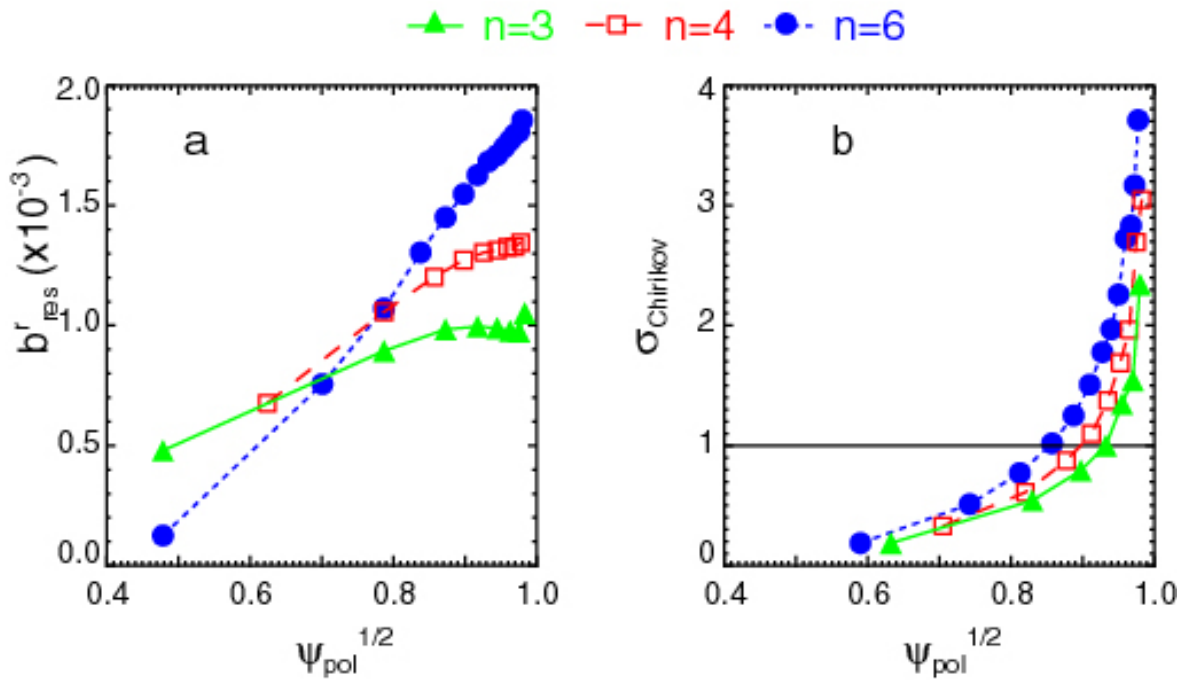

**Figure 4** Calculations in the vacuum approximation of a) the normalised resonant component of the applied field ($b^r_{res}$) and b) the Chirikov parameter profile produced for shots with the RMPs in n=3,4 and 6 configurations.

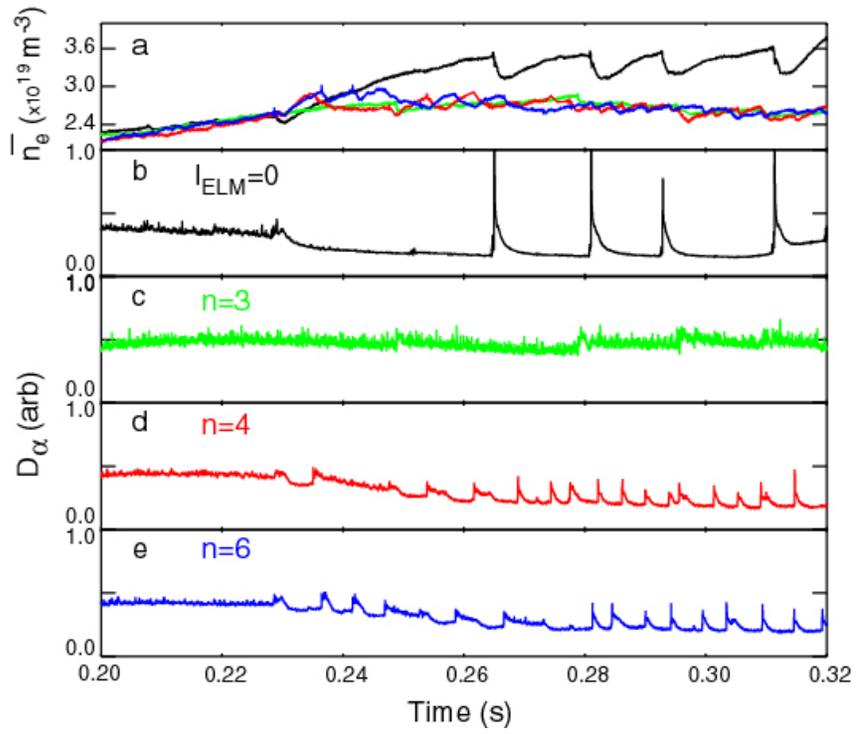

**Figure 5** Time traces of a) line average density ($\bar{n_e}$) and the target $D_\alpha$ intensity for a lower SND shot with the RMPs having a constant value of b) $I_{ELM} = 0kAt$, c) $I_{ELM} = 4.0kAt$ in n=3, d) $I_{ELM} = 5.6kAt$ in n=4 and e) $I_{ELM} = 5.6$ kAt in n=6 configuration.

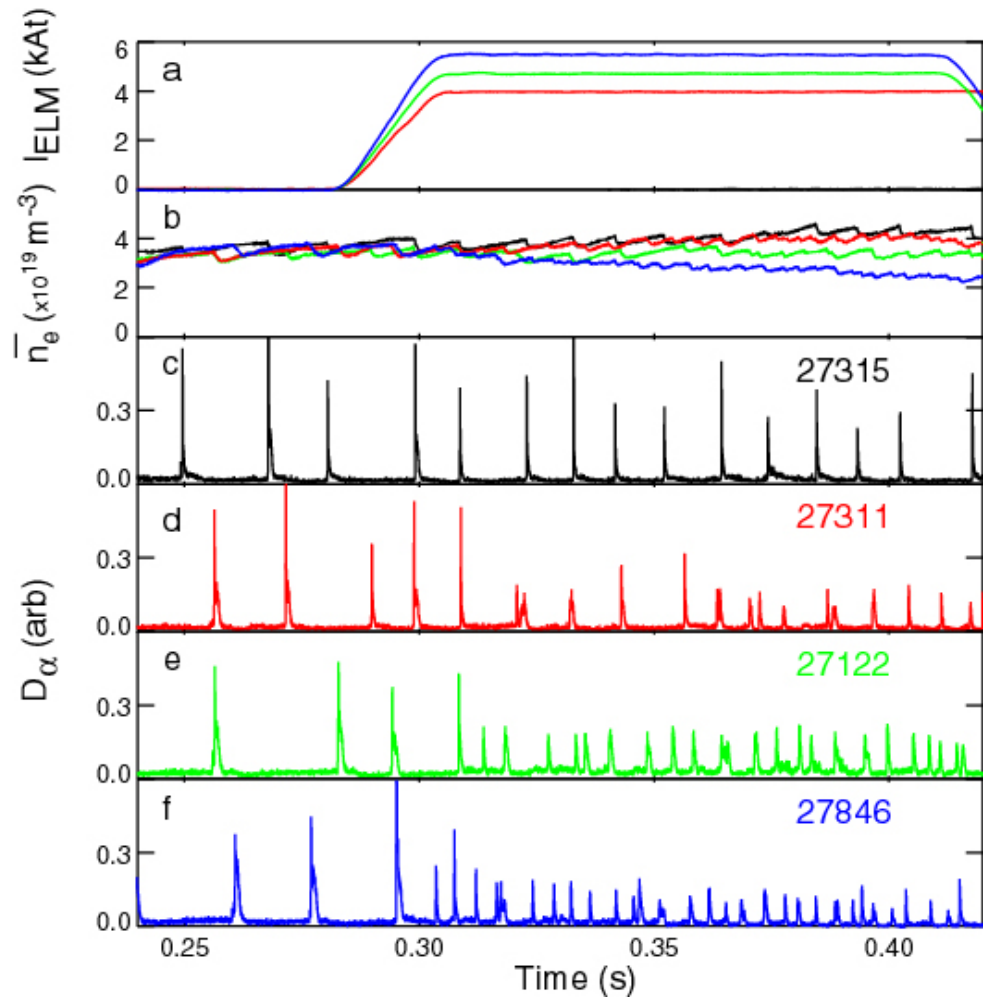

**Figure 6**  Time traces of a) the current in the ELM coils ($I_{ELM}$) b) line average density ($\bar{n}_e$), and the target $D_\alpha$ intensity for a lower SND shot with the RMPs in an n=4 configuration with c) $I_{ELM}$ = 0, b) $I_{ELM}$ = 4.0 kAt, c) $I_{ELM}$ = 4.8 kAt and d) $I_{ELM}$ = 5.6 kAt.

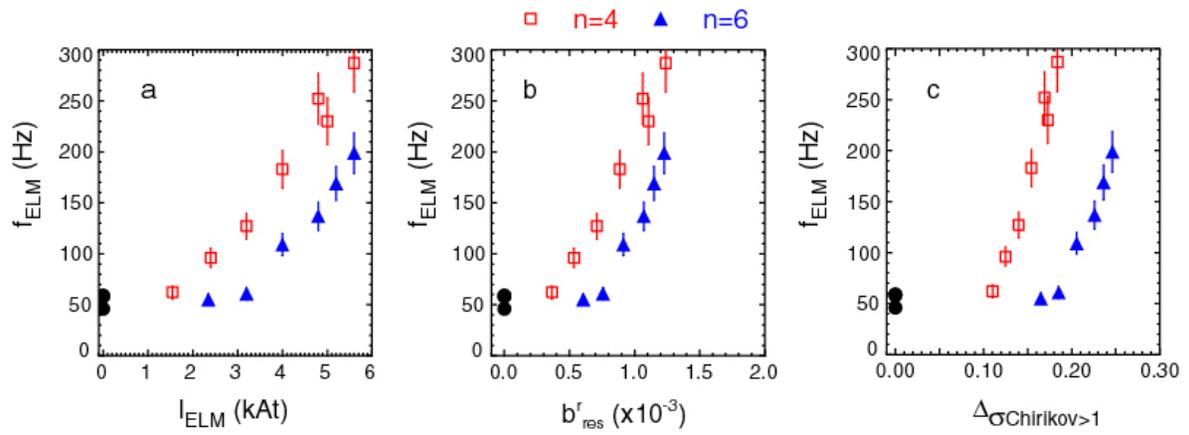

**Figure 7** ELM frequency ($f_{ELM}$) as a function of a) current in the ELM coils ($I_{ELM}$), b) maximum resonant component of the applied field ($b^r_{res}$) and c) the width of the region in $\Psi_{pol}$ for which the Chirikov parameter is greater than 1 ($\Delta_{\sigma Chirikov>1}$) for shots with the RMPs in an n=4 (open squares) and n=6 (closed triangles).

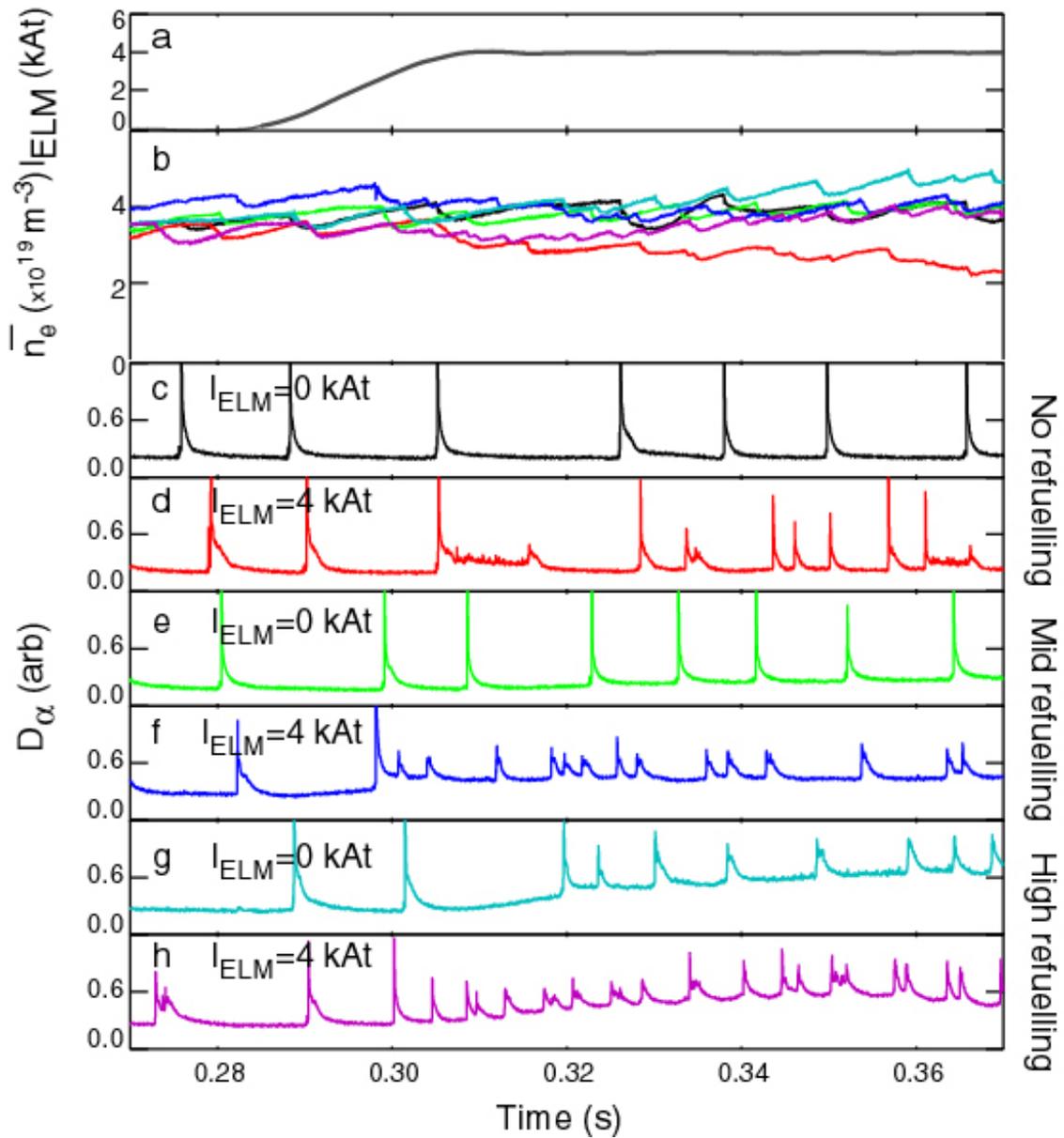

**Figure 8** Time traces of a) the current in the ELM coils ($I_{ELM}$) b) line average density ($\bar{n_e}$), and the target $D_\alpha$ intensity for a lower SND shot with the RMPs in an n=4 configuration with c) $I_{ELM}$ = 0kAt and d) $I_{ELM}$=4 kAt with no refuelling, e) $I_{ELM}$ = 0kAt and f) $I_{ELM}$=4 kAt with a refuelling rate of $4.6\times10^{21}$ D$_2$ s$^{-1}$ (mid refuelling) and g) $I_{ELM}$ = 0kAt and h) $I_{ELM}$=4 kAt with a refuelling rate of $8.0\times10^{21}$ D$_2$ s$^{-1}$ (high refuelling).

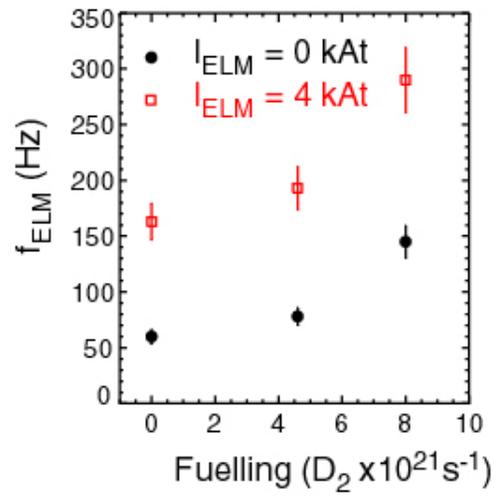

**Figure 9** Comparison of ELM frequency ($f_{ELM}$) as a function of fuelling rate for shots with $I_{ELM}$ = 0 kAt (solid circles) and $I_{ELM}$=4kAt (open squares) in the RMPs in an n=4 configuration.

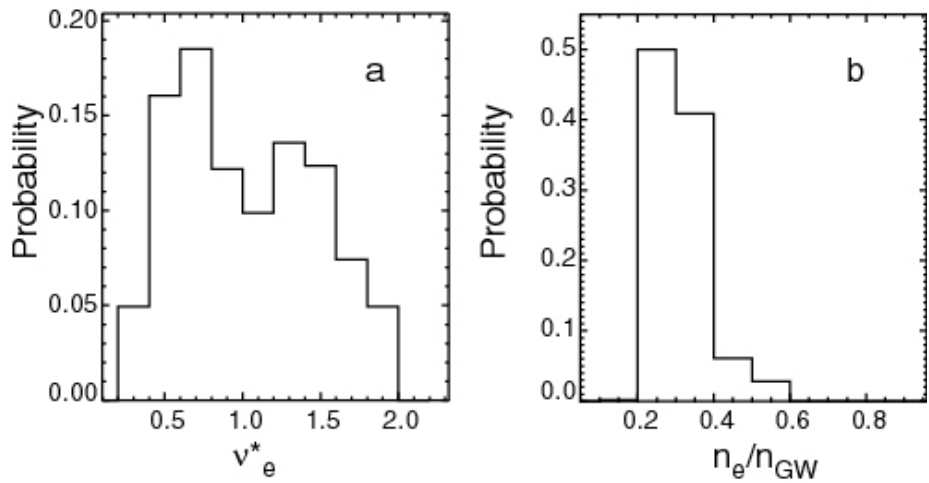

**Figure 10** Probability distribution of a) the edge pedestal edge collisionality ($v^*_e$) and b) the line averaged density as a fraction of the Greenwald density ($n_e/n_{GW}$).

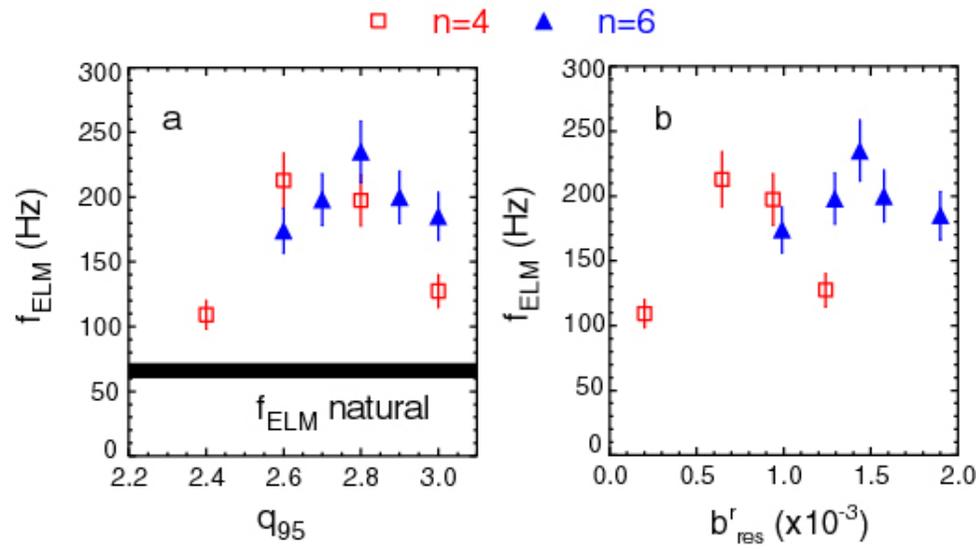

**Figure 11** ELM frequency ($f_{ELM}$) as a function of a) $q_{95}$ and b) maximum resonant component of the applied field ($b^r_{res}$) for shots with $I_{ELM}$ =4.0 kAt in the RMPs in an n=4 (open squares) and $I_{ELM}$ =5.6 kAt in an n=6 (closed triangles) configuration.

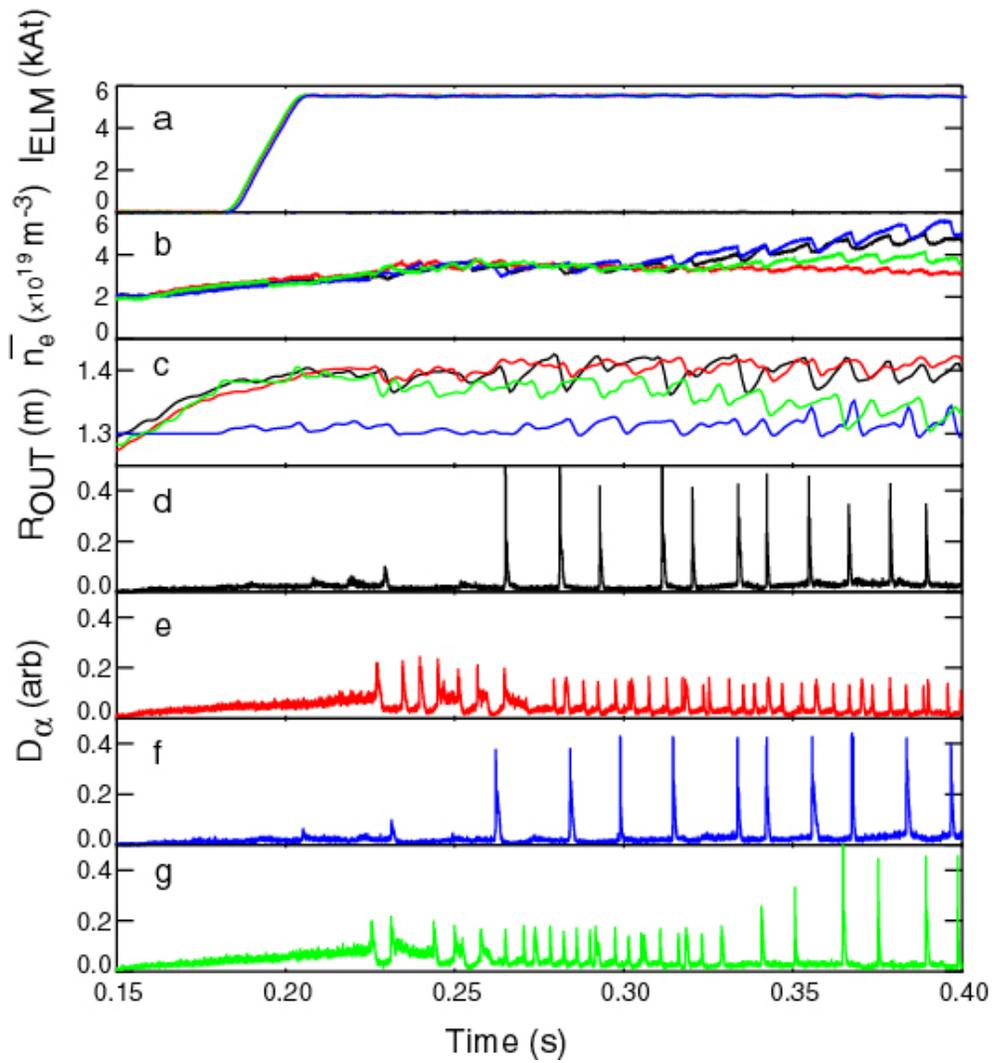

Figure 12 Time traces of a) the current in the ELM coils ($I_{ELM}$) b) line average density ($\bar{n_e}$), c) the outer radius of the plasma ($R_{OUT}$) and the target $D_\alpha$ intensity for a lower SND shot with the RMPs in an n=6 configuration with d) $I_{ELM} = 0kAt$ and e) $I_{ELM}=5.6$ kAt with $R_{OUT} = 1.4$ m, f) $I_{ELM} = 5.6kAt$ and $R_{OUT}=1.3$ m and g) $I_{ELM}=5.6$ kAt with $R_{OUT}$ decreasing from 1.4 m to 1.3m.

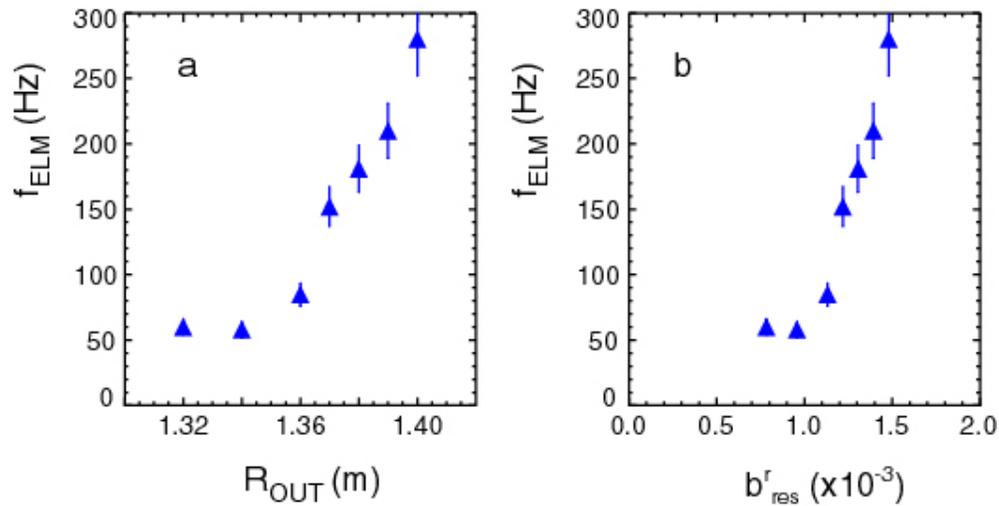

**Figure 13** ELM frequency ($f_{ELM}$) as a function of a) outer radius of the plasma ($R_{OUT}$) and b) maximum resonant component of the applied field ($b^r_{res}$) for shots with $I_{ELM} = 5.6$ kAt in the RMPs in an n=6 configuration.

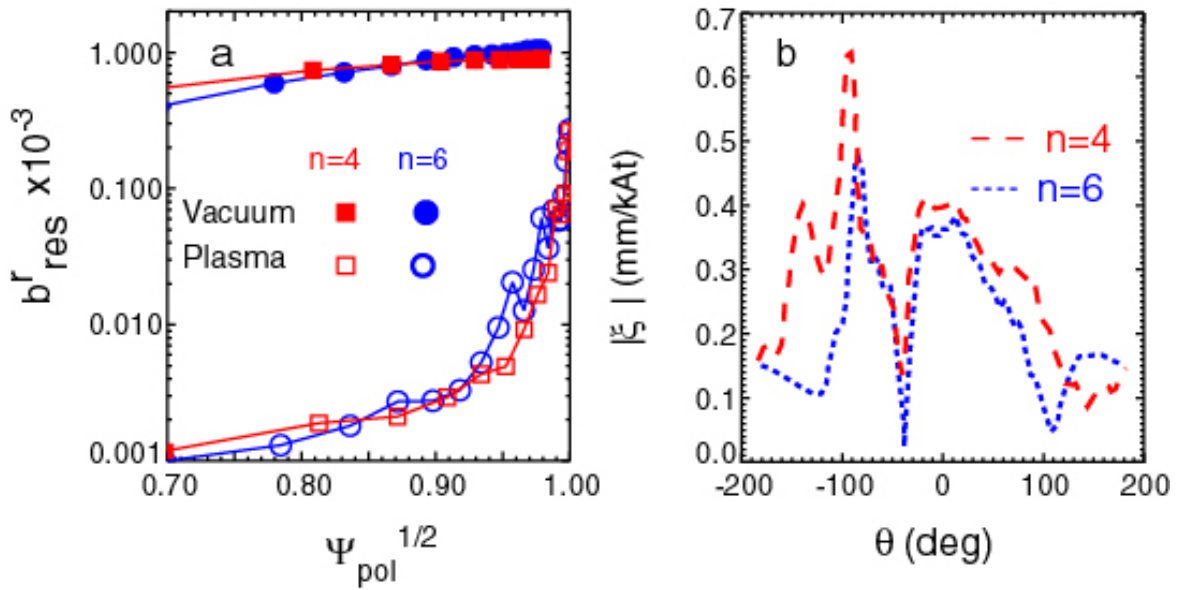

**Figure 14** a) Calculated profiles of the normalised resonant component of the applied field ($b^r_{res}$) produced with 5.6 kAt in the ELM coils in an n=4 (squares) and n=6 (circles) configuration using the vacuum approximation (solid) or taking into account the plasma response (open). b) The amplitude of the normal displacement of the plasma surface ($\xi$) computed by MARS-F as a function of poloidal angle ($\theta$) for RMPs in an n=4 (dashed) or n=6 (dotted) configuration.

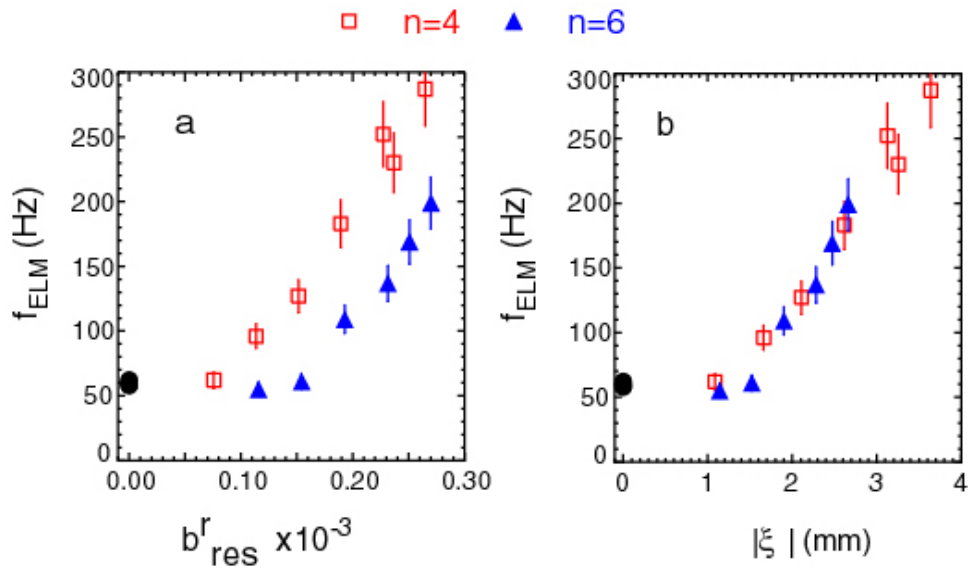

**Figure 15** Results from a scan of $I_{ELM}$: ELM frequency ($f_{ELM}$) as a function of a) maximum resonant component of the applied field ($b^r_{res}$) calculated taking into account the plasma response and b) the amplitude of the normal displacement of the plasma surface ($\xi$) at the X-point for shots with the RMPs in an n=4 (open squares) and n=6 (closed triangles) configuration.

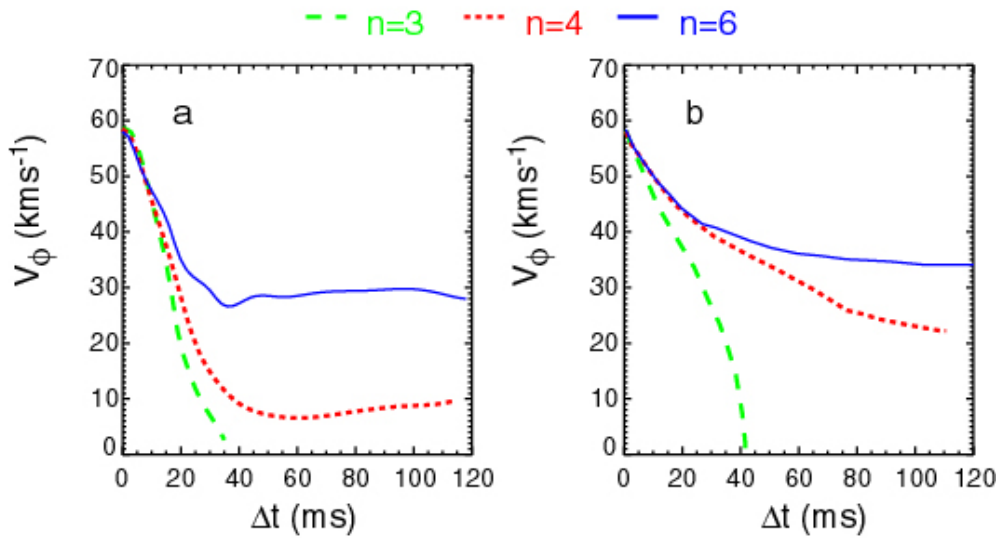

**Figure 16** a) The experimentally measured core toroidal rotation velocity as a function of time after which the RMPs reached flat top ($\Delta t$) and b) the results from the MARS-Q code simulations for shots with RMPs in n=3 (dashed) ,4 (dotted) and 6 (solid) configurations.

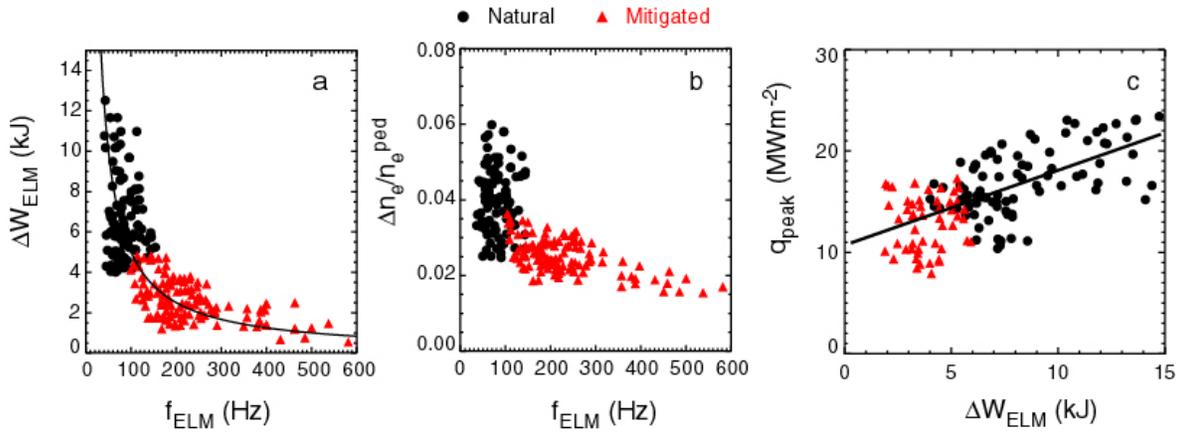

**Figure 17** a) ELM energy loss ($\Delta W_{ELM}$) as a function of ELM frequency ($f_{ELM}$) b) ELM particle loss expressed as a fraction of the pedestal density ($\Delta n_e/n_e^{ped}$) and c) maximum peak heat flux during an ELM at the low field side divertor as a function of $f_{ELM}$ for natural ($I_{ELM}=0$ kAt) and mitigated ELMs.

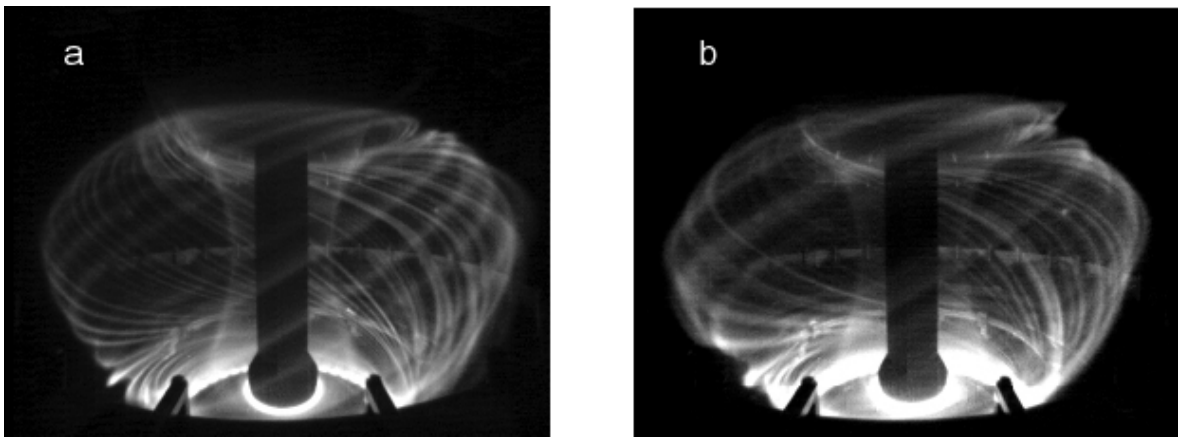

**Figure 18** Visible images obtained during an ELM for a) a natural ELM and b) a mitigated ELM produced by RMPs in an n=6 configuration with $I_{ELM}=5.6$ kAt.

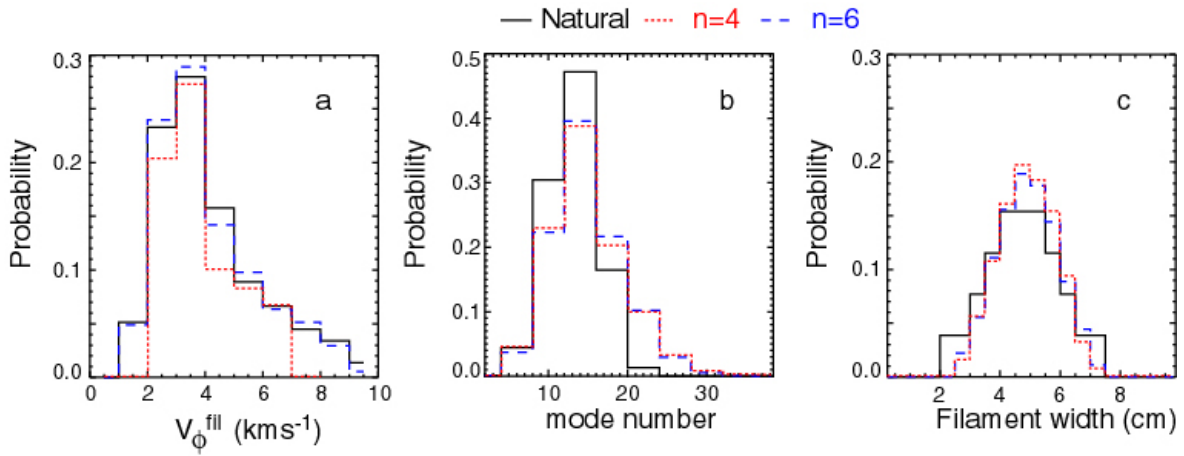

**Figure 19** Probability distribution of a) the toroidal velocity of the filaments ($V_\phi$), b) the toroidal mode number and c) the filament width for natural ELMs (solid) and mitigated ELMs obtained with the RMPs in an n=4 (dotted) or n=6 (dashed) configuration.

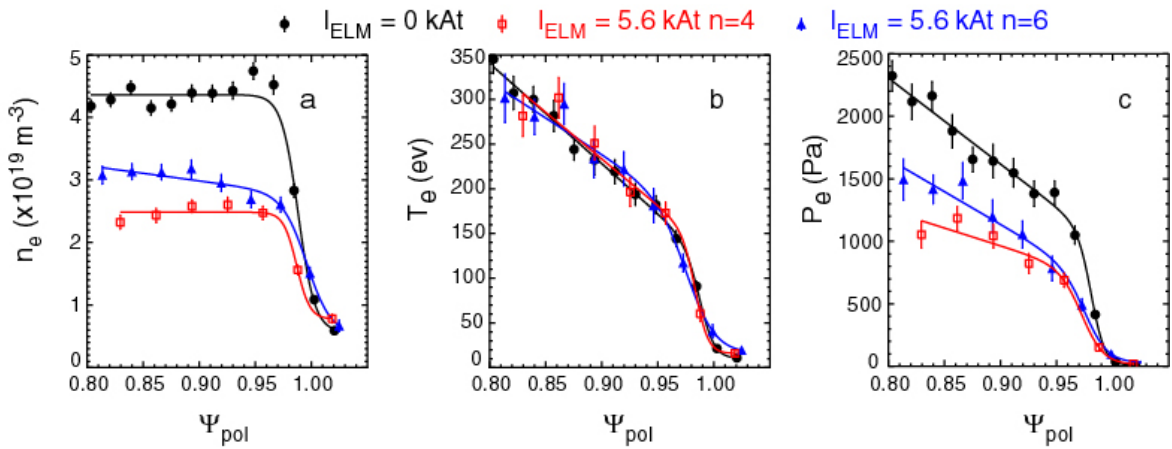

**Figure 20** Comparison of the profiles of electron a) density ($n_e$) , b) temperature ($T_e$) and c) pressure ($P_e$) in normalised poloidal flux ($\Psi_{pol}$) space for shots with $I_{ELM} = 0$ kAt (closed circle) and $I_{ELM} = 5.6$ kAt with the RMPs in an n=4 (open square) and n=6 (closed triangle) configurations.

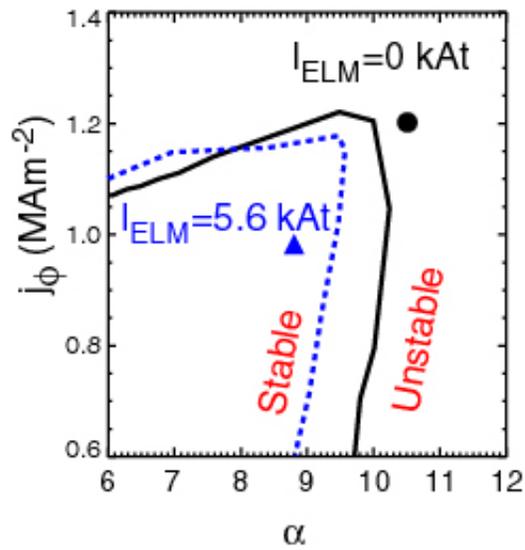

**Figure 21** Edge stability diagram plots of edge current density ($j_\phi$) versus normalised pressure gradient ($\alpha$) calculated for a shot with $I_{ELM} = 0$ kAt (solid line) and for a shot with $I_{ELM} = 5.6$kAT in the RMPs in an n=6 configuration (dashed line). The solid circle and triangle represent the experimental points for the $I_{ELM} = 0$ and 5.6 kAt cases respectively.

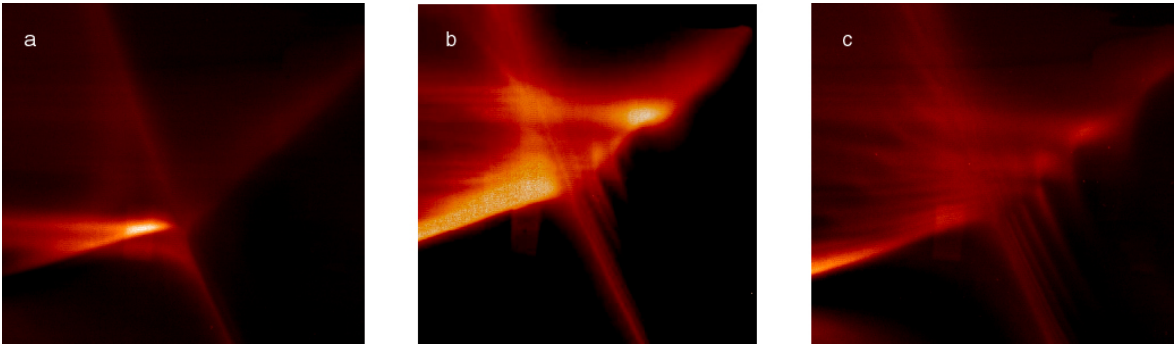

**Figure 22** False colour images of the He II emission from the X-point region captured during an Inter-ELM period of an H-mode a) without RMPs and with $I_{ELM} = 5.6$kAT in the RMPs in b) an n=6 and c) an n=4 configuration.

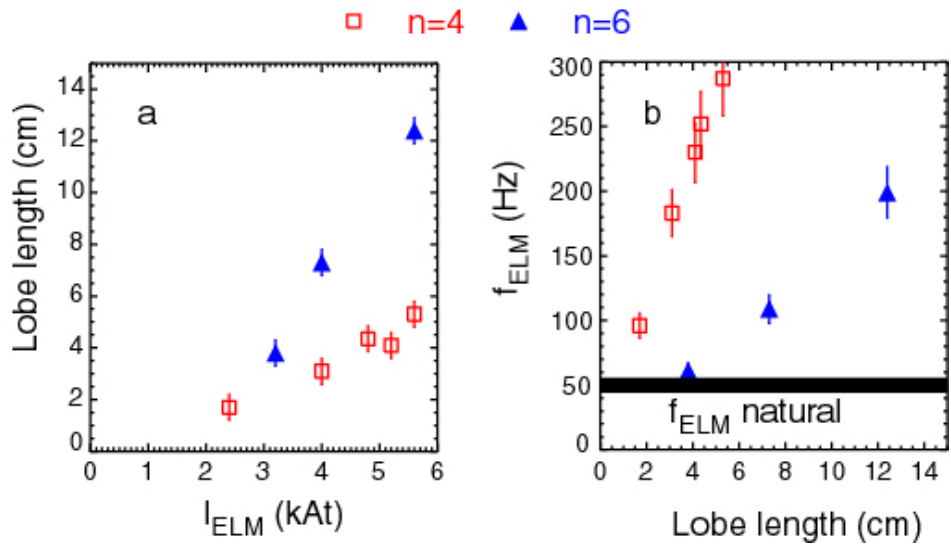

**Figure 23** a) Lobe length versus current in the ELM coils ($I_{ELM}$) and b) ELM frequency ($f_{ELM}$) versus lobe length for the RMPs in an n=4 (open squares) and n=6 (closed triangles) configuration.